\documentclass[onecolumn,noshowpacs,preprintnumbers,amsmath,amssymb,prb,bibtex]{revtex4}
\usepackage{graphicx}
\usepackage{color}
\usepackage{graphicx}
\usepackage{dcolumn}
\usepackage{bm}
\usepackage{bbm}
\usepackage{setspace}
\usepackage{braket}
\usepackage{multirow}
\usepackage{enumitem}
\usepackage[hang,small,bf]{caption}
\usepackage[subrefformat=parens]{subcaption}
\usepackage{here}

\newcommand{\al}{\alpha}
\newcommand{\be}{\beta}

\newcommand{\si}{\sigma}
\newcommand{\ta}{\tau}

\newcommand{\da}{\dagger}
\newcommand{\la}{\lambda}

\newcommand{\epl}{\epsilon}

\newcommand{\ahat}{\hat{a}}

\newcommand{\atil}{\tilde{a}}

\newcommand{\circn}[1]{$^{\text{circ/}#1}$}

\begin{document}

\title{A Jastrow-type decomposition in quantum chemistry for low-depth quantum circuits}

\date{September 26, 2019}

\author{Yuta Matsuzawa}
\affiliation{Department of Chemistry, Graduate School of Science, Kyoto University, Kitashirakawa Oiwake-cho, Sakyo-ku Kyoto, 606-8502, Japan}

\author{Yuki Kurashige\footnote{Electronic mail: kura@kuchem.kyoto-u.ac.jp.}}
\affiliation{Department of Chemistry, Graduate School of Science, Kyoto University, Kitashirakawa Oiwake-cho, Sakyo-ku Kyoto, 606-8502, Japan}

\begin{abstract}
We propose an efficient ${\cal O}(N^2)$-parameter ansatz that consists
of a sequence of exponential operators, each of which is a unitary
variant of Neuscamman's cluster Jastrow operator. The ansatz can also
be derived as a decomposition of T$_2$ amplitudes of the unitary
coupled cluster with generalized singles and doubles, which gives a
near full-CI energy, and reproduces it by extending the
exponential operator sequence. Because the cluster Jastrow operators
are expressed by a product of number operators and the derived Pauli
operator products, namely the Jordan-Wigner strings, are all
commutative, it does not require the Trotter approximation to
implement to a quantum circuit and should be a good candidate for the
variational quantum eigensolver algorithm by a near-term quantum
computer.
The accuracy of the ansatz was examined for dissociation of a nitrogen
dimer, and compared with other existing ${\cal O}(N^2)$-parameter
ansatzs. Not only the original ansatzs defined in the
second-quantization form but also their Trotterized variants, in which
the cluster amplitudes are optimized to minimize the energy obtained
with a few, typically single, Trotter steps, were examined by quantum
circuit simulators.
\end{abstract}

\maketitle
\setstretch{1.1}

\section{Introduction}
Quantum chemistry is gaining more attention as a promising field of
application of quantum computers because of its high affinity to the
qubit representations and universal gate
operations.\cite{lanyon2010white, wecker2014troyer, mcardle2018yuan-1,
  cao2018aspuru-guzik, berry2019babbush} An obvious advantage of
quantum computers on quantum chemical calculations is that
occupation-number vectors of the second quantization can be naturally
mapped to the qubit representation by the Jordan-Wigner or
Bravyi-Kitaev transformations, and arbitrary state in the vector space
and unitary transformation are expressed by the qubits and quantum
gates, while the space is usually truncated for the case of classical
computers because the dimension of the vector space, namely the
full-CI space, grows exponentially with the number of one-particle
basis.

While quantum algorithms for numerical linear algebra is currently
being developed,\cite{motta2019chan, huggins2019whaley,
  parrish2019mcmahon} so far there are two main approaches for
computing eigenvalues of the second-quantized electronic
Hamiltonian. The first developed method is the phase estimation
algorithm (PEA),\cite{kitaev1995kitaev-1,
  aspuru-guzik2005aspuru-guzik} in which an approximated eigenstate
prepared on quantum registers is propagated with the Hamiltonian to
encode the eigenvalue into the relative phases of binary states of
ancilla qubits and the phase is decoded by a fractional binary
expansion $via$ the inverse quantum Fourier transformation (QFT).
In theory, a desired precision $\epl$ can be achieved with a cost of
${\cal O}(1/\epl)$, but in reality it largely increases the demand for
coherence time, which is severe for near-term quantum computers
without error corrections.
There have been many attempts to mitigate the requirements, such as
the iterative PEA\cite{dobsicek2007wendin, omalley2016martinis} which
does not require the QFT.
The second method is the variational quantum eigensolvers
(VQE)\cite{peruzzo2014obrien, mcclean2016aspuru-guzik} in which a
quantum computer is used only for evaluating energy expectation values
of trial wavefunctions that are defined by a set of parameters, which
are optimized by a classical computer. One of the advantages of the
VQE over the PEA is that once a trial wavefunction is prepared, the
measurements only require simple single qubit rotations, which should
greatly reduce the coherence time requirements, ideally ${\cal
  O}(1)$.\cite{mcclean2014aspuru-guzik} In addition, the recent
developments of the quantum algorithms\cite{mcardle2018yuan,
  endo2019benjamin, mitarai2018fujii, mitarai2019mizukami} for the
evaluation of the derivatives with respect to the parameters in the
exponents should greatly accelerate the variational optimization.
The VQE is therefore a promising candidate algorithm in the noisy
intermediate scale quantum computers (NISQ)
era,\cite{preskill2018preskill} though the required times of
measurements increases as ${\cal O}(1/\epl^2)$ to achieve a desired
precision $\epl$.

It is therefore important to develop an efficient wavefunction ansatz
that requires only a small number of quantum gates to prepare on a
quantum circuit. The widely-used unitary coupled cluster with singles
and doubles (uCCSD) ansatz\cite{bartlett1989noga, yanai2006chan,
  harsha2018scuseria} has ${\cal O}(N^4)$ variational parameters and
the depth should grow as ${\cal O}(N^3)$ at least. A number of ${\cal
  O}(N^2)$-parameter ansatz have been developed, such as the low-depth
ansatz,\cite{dallaire-demers2018aspuru-guzik} and low-rank
approximation,\cite{motta2018chan} and recently, k-fold products of
unitary pair coupled cluster with generalized singles and doubles
(UpCCGSD).\cite{lee2019whaley}
Another obstacle to implement exponential operator ansatzs to quantum
circuits is that all the Pauli operator products produced by the
Jordan-Wigner or Bravyi-Kitaev transformations of the second quantized
operators in an exponent need to be split to a product of the
exponential operators, each of which has only single Pauli product in
the exponenet. It means many Trotter steps should be performed before
the measurement if one wants to compute the energy expectation value
accurately as provided by the definition written in the second
quantized operators. One way to work around this problem is to adopt
hardware oriented ansatzs, such as the hardware efficient
ansatz\cite{kandala2017gambetta} or an incremental adaptive
scheme.\cite{grimsley2019mayhall}

In this study, we present an efficient ${\cal O}(N^2)$-parameter
ansatz for the VQE algorithm termed $k$-uCJ that consists of repeated
$k$-times multiplication of a unitary variant of the cluster Jastrow
(CJ) exponential operator,\cite{neuscamman2013neuscamman-1,
  neuscamman2016neuscamman} and can reproduce the generalized unitary
coupled cluster (uCCGSD) ansatz\cite{nakatsuji2000nakatsuji,
  nooijen2000nooijen, lee2019whaley} by increasing $k$, which can give
a near full-CI energy.

\section{\label{sec:theory}Theory}
In this section, matrices can be real or complex. For clarification,
they are denoted by italic and calligraphy letters, e.g. $K$ and ${\cal
  K}$, respectively.

\subsection{\label{subsec:theory_XXX} Unitarization of the CJ ansatz}
An efficient parametrization for T$_2$ amplitudes of generalized CC,
named as cluster Jastrow (CJ) ansatz
\begin{align}
T_{p_\si q_\si r_\ta s_\ta}
\simeq T^\text{\tiny CJ}_{p_\si q_\si r_\ta s_\ta}
= \sum_{jl} U_{pj} U_{qj} J^{(\si \ta)}_{jl} U_{rl} U_{sl},
\end{align}
was introduced by Neuscamman\cite{neuscamman2013neuscamman-1,
  neuscamman2016neuscamman} in his cluster Jastrow anti-symmetric
geminal power (CJAGP) ansatz. Simultaneous optimization of the
parameters $U$, $J$ and the AGP reference function with a variational
Monte Carlo method provided an efficient and accurate description of
both weak and strong electron
correlation.\cite{neuscamman2012neuscamman, neuscamman2013neuscamman,
  neuscamman2013neuscamman-1, neuscamman2016neuscamman}
In the CJ ansatz, the 4-index cluster operator $\hat{T}^\text{\tiny
  CJ}$ is decomposed to two 2-index operators
\begin{align}
e^{\hat{T}^\text{\tiny CJ}}
&= e^{\hat{K}} e^{\hat{J}} e^{-\hat{K}}\label{eq:CJ}
\end{align}
where
\begin{align}
\hat{T}^\text{\tiny CJ}
&= \sum_{pqrs,\si\ta} T^\text{\tiny CJ}_{p_\si q_\si r_\ta s_\ta}
  a^\da_{p\si} a_{q\si}  a^\da_{r\ta} a_{s\ta}
\\
\hat{J} &= \sum_{jl,\si\ta} J^{\si \ta}_{jl} a^\da_{j\si} a_{j\si}  a^\da_{l\ta} a_{l\ta}.\label{eq:matJ}
\\
\hat{K} &=\sum_{p<q,\si} K_{pq} (a^\da_{p\si} a_{q\si} - a^\da_{q\si} a_{p\si})\label{eq:matK}
\end{align}
While $e^{\hat{K}}$ is an orbital rotation operator and easy to map to
quantum gates, $\hat{J}$ is symmetric with respect to permutation of
the indices $j$ and $l$ and thus $e^{\hat{J}}$ cannot be unitary if
one restricts $J$ to real matrices.
The matrix $J$ needs to be replaced by pure imaginary matrices ${\cal
  J}$ in order to make $e^{\hat{J}}$ unitary operator, and we
introduce a unitary variant of the CJ ansatz, uCJ ansatz, by replacing
$\hat{K}$ and $\hat{J}$ in Eq.(\ref{eq:CJ}) by
\begin{align}
\hat{K} &=\sum_{pq,\si} {\cal K}_{pq} a^\da_{p\si} a_{q\si}\\
\hat{J} &= \sum_{jl,\si\ta} {\cal J}^{(\si\ta)}_{jl} a^\da_{j\si} a_{j\si}  a^\da_{l\ta} a_{l\ta}.
\end{align}
where ${\cal K}$ is a complex anti-Hermite matrix and ${\cal
  J}^{\al\al}($=${\cal J}^{\be\be})$, ${\cal J}^{\al\be}($=${\cal
  J}^{\be\al})$ are pure imaginary symmetric matrices with respect to
the indices $k$ and $l$. Note that when we restricted ${\cal K}$ to
real anti-symmetric matrices, energy expectation values by the ansatz
never fell below that of reference wavefunctions. The cluster
amplitudes of uCJ are therefore generally complex numbers
\begin{align}
T_{p_\si q_\si r_\ta s_\ta}
\simeq
{\cal T}^\text{\tiny uCJ}_{p_\si q_\si r_\ta s_\ta}
&= \sum_{jl} {\cal U}_{pj} {\cal U}^*_{qj} {\cal J}^{(\si\ta)}_{jl} {\cal U}_{rl} {\cal U}^*_{sl}.\label{eq:uCJ}
\end{align}
It is possible to restrict the amplitudes to real numbers by adding
its complex conjugate ${\cal T}^\text{\tiny uCJ}_{p_\si q_\si r_\ta
  s_\ta} + {\cal T}^{\text{\tiny uCJ}*}_{p_\si q_\si r_\ta s_\ta}$,
but we found that the variational energy is not different from that
with Eq.(\ref{eq:uCJ}).

\subsection{\label{subsec:theory_amp} Extention of uCJ ansatz}
To recover the uCCGSD limit in Eq.~(\ref{eq:uCJ}), it is natural to
extend the uCJ ansatz by introducing an extra index $x$ to ${\cal J}$
and ${\cal K}$ as
\begin{align}
{\cal T}_{p_\si q_\si r_\ta s_\ta}
\simeq \sum_{x=1}^{k} \sum_{jl} {\cal U}^{x}_{pj} {\cal U}^{x*}_{qj}
            {\cal J}^{(\si\ta)x}_{jl} {\cal U}^{x}_{rl} {\cal U}^{x*}_{sl}.\label{eq:k-uCJ}
\end{align}
It is closely related to the low-rank approximation to the CCSD
amplitudes recently introduced by Motta and
co-workers,\cite{motta2018chan} and originally by Peng and
Kowalski\cite{peng2017kowalski} to 4-index two-electron repulsion
integrals as will be described in Appendix.
The low-rank decomposition can be easily extended to uCCGD cluster
operator as
\begin{align}
\hat{\ta}^\text{\tiny uCCG}
&
= \hat{T}^\text{\tiny CCG} - \hat{T}^{\text{\tiny CCG}\da}
= \sum_{pqrs}\tau^\text{\tiny uCCG}_{pqrs} \ahat_p^\da  \ahat_q \ahat_r^\da  \ahat_s
\nonumber\\
&= \frac{1}{2}\sum_{x}\Bigg\{
 (\Sigma_{pq} {\cal L}^{x+}_{pq} \ahat_p^\da  \ahat_q)^2
+(\Sigma_{pq} {\cal L}^{x-}_{pq} \ahat_p^\da  \ahat_q)^2
\Bigg\}\nonumber\\
&= \frac{1}{2}\sum_{x}\Bigg\{
\Big(\sum_{pq} (\Sigma_j {\cal U}^{x+}_{pj} \la^{x+}_j {\cal U}^{x+*}_{qj}) \ahat_p^\da  \ahat_q
\Big)^2
\nonumber\\
&\phantom{\frac{1}{2}\sum_{x}\Bigg\{}+
\Big(\sum_{pq} (\Sigma_j {\cal U}^{x-}_{pj} \la^{x-}_j {\cal U}^{x-*}_{qj}) \ahat_p^\da  \ahat_q
\Big)^2
\Bigg\}
\end{align}
thus
\begin{align}
\tau^\text{\tiny uCCG}_{pqrs}
&= \frac{1}{2}\sum^{k}_{x} \sum_{jl}
\Bigg\{
{\cal U}^{x+}_{pj} {\cal U}^{x+*}_{qj} (\la^{x+}_j \la^{x+}_l)
{\cal U}^{x+}_{rl} {\cal U}^{x+*}_{sl}
\nonumber\\
&\phantom{\frac{1}{2}\sum^{k}_{x} \sum_{jl}\Bigg\{}+
{\cal U}^{x-}_{pj} {\cal U}^{x-*}_{qj} (\la^{x-}_j \la^{x-}_l)
{\cal U}^{x-}_{rl} {\cal U}^{x-*}_{sl}
\Bigg\}\label{eq:low-rank}
\end{align}
where ${\cal L}^\pm$, ${\cal U}^\pm$, and $\la^\pm$ are
obtained by SVD
\begin{align}
T_{pq,rs}
&= \sum_{x} V_{pq,x} \si_{x} V_{rs,x}
\nonumber\\
&= \sum_{x} (\sqrt{\si_{x}} V_{pq,x}) (\sqrt{\si_{x}} V_{rs,x})\nonumber\\
&=
\sum_{x} \frac{\si_{x}}{|\si_{x}|} L^{x}_{pq} L^{x}_{rs}
\end{align}
and SVD again
\begin{align}
{\cal L}^{x\pm}_{pq}
&= \sqrt{\frac{\si_{x}}{|\si_{x}|}} (L^{x}_{pq} \pm i L^{x}_{qp})\label{eq:L_pm}\\
&= \sum_j {\cal U}^{x\pm}_{pj} \la^{x\pm}_j {\cal U}^{x\pm *}_{qj}.
\end{align}
Because $L^{x}$ are not generally symmetric matrices,
Eq.~(\ref{eq:L_pm}) is needed to make ${\cal L}^{x\pm}$ normal
matrices and to ensure the existence of unitary matrices ${\cal
  U}^{x\pm}$.
The decomposition by Eq.~(\ref{eq:k-uCJ}) can give the same
$\tau^\text{\tiny uCCG}$ amplitude by setting ${\cal J}^{x}_{jl} =
\la^{x+}_j \la^{x+}_l$, ${\cal U}^{x}_{pj} = {\cal U}^{x+}_{pj}$ and
adding its complex conjugate, rather is much more flexible than
Eq.(\ref{eq:low-rank}) because ${\cal J}^{x}$ are not restricted to
rank-one matrices. The convergence to $\tau^\text{\tiny uCCG}$
amplitude should be much faster, i.e.\ more accurate with the same
number of terms, $k$, for truncating the summation over
$x$. Hereafter, They are denoted by the SVD($k$) and uCJ($k$)
decompositions, respectively.

Instead of the single exponent form
$
e^{\sum^k_{x=1} {\hat{\ta}}^\text{\tiny uCJ}_{(x)}}
$
,
we adopt a product form of the uCJ exponential operators
\begin{align}
&e^{{\hat{\ta}}^\text{\tiny uCJ}_{(k)}} \cdots\;\;
 e^{{\hat{\ta}}^\text{\tiny uCJ}_{(2)}}
 e^{{\hat{\ta}}^\text{\tiny uCJ}_{(1)}}\nonumber\\
&=
e^{-\hat{K}_{k}} e^{\hat{J}_{k}} e^{\hat{K}_{k}} \cdots\;\;
e^{-\hat{K}_{2}} e^{\hat{J}_{2}} e^{\hat{K}_{2}}
e^{-\hat{K}_{1}} e^{\hat{J}_{1}} e^{\hat{K}_{1}}\label{eq:product_t2}
\end{align}
since the latter is more suitable for quantum computation. In this
study, it is termed as $k$-uCJ ansatz since this splitting exponential
operator form is analogous to k-fold products of ($k$-)UpCCGSD ansatz
recently introduced by Lee and co-workers,\cite{lee2019whaley} in
which a product of exponential operators of the pair coupled cluster
doubles\cite{stein2014scuseria, limacher2013bultinck,
  boguslawski2014vanneck, boguslawski2015ayers, henderson2015scuseria}
with the generalized singles.
Note that the product form and the single exponent form are not
equivalent because the uCJ operators with different $x$ are not
commutative, but we observed that these two forms gave very close
variational energies if all the parameter are simultaneously optimized
based on each ansatz.
%


\subsection{\label{subsec:theory_qcomp}
Some consideratins for implementation of the exponential ansatzs on
quantum circuits}
In principle, any unitary operation can be represented by quantum
gates on universal quantum computer, but so far most of the quantum
algorithms for quantum chemistry rely on the mapping of an exponential
of a product of Pauli operators to a set of quantum gates consisting
of CNOT and single qubit gates.\cite{nielsen2010chuang} Those Pauli
operators are produced by the Jordan-Wigner or Bravyi-Kitaev
transformations of the second-quantized operators of quantum
chemistry.\cite{seeley2012love}

For example, application of the elementary creation and annihilation
operators are represented by the Pauli-$X,Y,Z$ gates in the
Jordan-Wigner transformation as
\begin{align}
a_p^\dagger 
            &= \frac{1}{2}(\si^x_p \otimes \si^{z\;\rightarrow}_{p-1} - i\si^y_p \otimes \si^{z\;\rightarrow}_{p-1}) \\
a_p         
            &= \frac{1}{2}(\si^x_p \otimes \si^{z\;\rightarrow}_{p-1} + i\si^y_p \otimes \si^{z\;\rightarrow}_{p-1})
\end{align}
where $\si^{z\;\rightarrow}_{p-1} \equiv \si^z_p \otimes \si^z_{p-1}
\otimes \cdots \si^z_2 \otimes \si^z_1$, and thus an exponential
operator $e^{\hat{\ta}}$ is represented by an exponential of a summation
of Pauli operator products, $e^{i\sum_{I} \theta_I (\Pi_{\mu\in I}
  \si_\mu)}$. Note that the imaginary unit $i$ in the exponent
reflects unitarity of the cluster operator $\hat{\ta}$.
To compute an exponential operator $e^{\hat{\tau}}$ on a quantum
circuit accurately, one therefore needs to resort to the Trotter
decomposition, e.g.\
\begin{align}
e^{i\sum_{I} \theta_I (\prod_{\mu\in I} \si_\mu)}
\simeq \Big(\prod_{I} e^{i \theta_I n^{-1}\prod_{\mu\in I} \si_\mu}\Big)^{n},
\end{align}
because the Pauli operator products $\prod_{\mu\in I} \si_\mu$ are
usually not commutative. The many Trotter steps should increase the
depth of the circuit. One way to avoid this problem is to simply adopt
the sigle Trotter step form $\prod_{I} e^{i\theta_I \prod_{\mu\in I}
  \si_\mu}$, or $n$-step form, as an alternative ansatz. In fact, it
was demonstrated that the single Trotter step ansatz of uCCSD gave
nearly identical variational energies to the original uCCSD ansatz for
a H$_2$ molecule.\cite{barkoutsos2018tavernelli}
Hereafter, the Trotter $n$-step ansatz is denoted by a superscript
`circ/$n$' to the corresponding original ansatz, e.g.\ the energy of
the $k$-CJF\circn{n} ansatzs are evaluated by performing the $n$-times
symmetric Trotter steps with the $k$-CJF ansatz, respectively.  In
fact, we found that not all exponential ansatz are compatible with
their Trotter $n$-step ansatzs as will be shown in
Sec.\ref{subsec:results_circ}.

Interestingly, the $k$-uCJ ansatz defined by Eq.~(\ref{eq:k-uCJ}) can
be implemented without the Trotter approximation by nature. Because
the operator $\hat{J}$ only have the number operators
\begin{align}
a^\da_p a_p a^\da_q a_q
&= \frac{1}{4} (1 - \si^z_p) (1 - \si^z_q)
,
\end{align}
which are written by $\si^z$ matrices, they are all commutative.  If
the orbital rotations $e^{\hat{K}}$ is implemented by using the
Givens rotations,\cite{wecker2015troyer, kivlichan2018babbush} one can
implement the $k$-uCJ to a quantum circuit without use of Trotter
approximation.

\section{\label{sec:results}Results and Discussion}
An important field of applications of quantum computation in quantum
chemistry is the multireference problem where rigorous algorithms with
polynomial cost are not well established though a lot of effort has
been devoted, such as {\it ab initio} density-matrix renormalization
group theory.\cite{white1999martin, chan2002head-gordon} We examined
the performance of various exponential ansatzs on the description of
triple bonds dissociation of N$_2$ molecule with the STO-6G basis
sets. The six Hartree-Fock canonical orbitals, HOMO$-2$ to LUMO$+2$,
were used to construct the Fock space represented by a quantum
register on a circuit simulator, i.e.\ the $\alpha$ and $\beta$
spin-orbital were assigned to twelve qubits. The ordering and
character of the six canonical orbitals remain unchanged between
$r=$1.0$-$2.4 \AA\ bond length.


\subsection{\label{subsec:results_amp}
Variational minimization without the Trotter approximation}

\begin{figure}
    \captionsetup{justification=raggedright}
    \hfill
    \begin{minipage}[t]{1.0\hsize}
      \subcaption{SVD and $k$-uCJ ($k$=1,2,3) ansatzs}
      \centering
      \includegraphics[width=8.5cm]{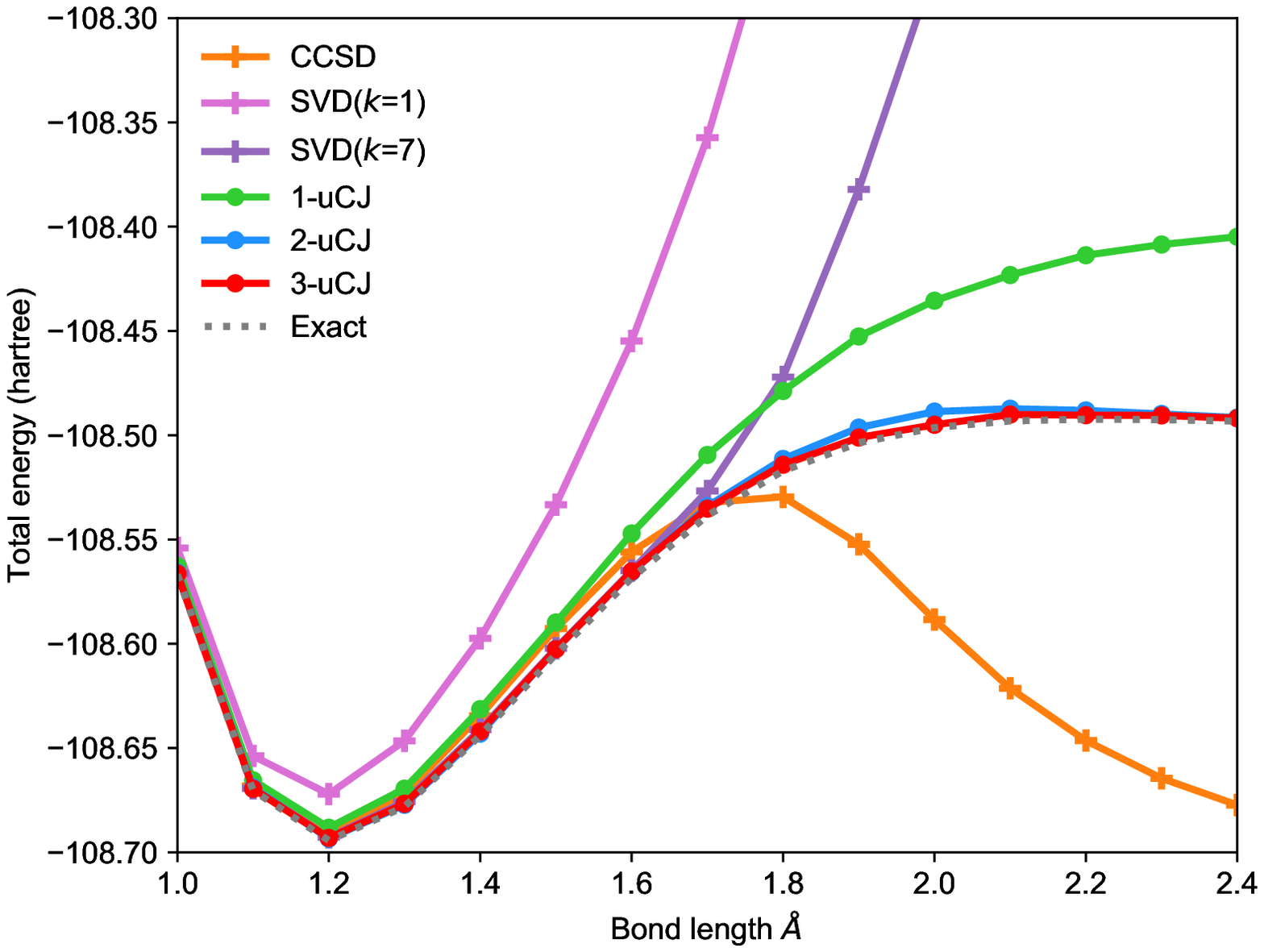}
      \label{fig:N2_SVD}
    \end{minipage}\\ 
    \hfill
    \begin{minipage}[t]{1.0\hsize}
      \subcaption{uCJ+AGP and $k$-UpCCGSD ($k$=1,2,3) ansatzs}
      \centering
      \includegraphics[width=8.5cm]{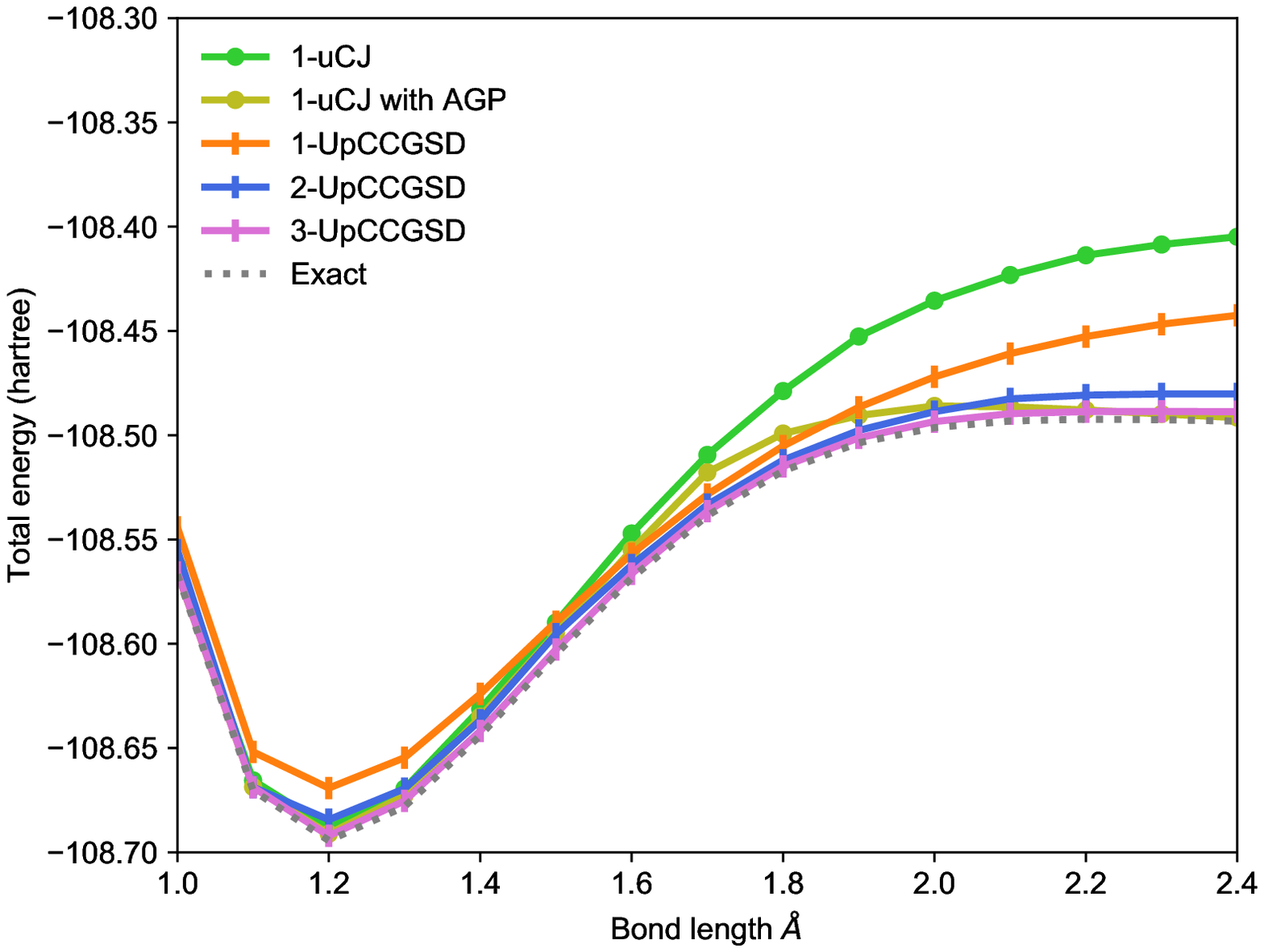}
      \label{fig:N2_UpCCGSD}
    \end{minipage}
    \caption{Potential energy curves for the N$_2$ molecule for the
      different ansatzs using the STO-6G basis set where the four lowest
      orbitals are frozen, i.e.\ 6 electrons are distributed to the
      other 6 orbitals.}
    \label{fig:N2_wo_trotter}
\end{figure}
%
%
First, we examined the accuracy of the $k$-uCJ ansatz and various
existing exponential ansatzs with the HF reference function. At this
point, the energy expectation values were evaluated not by using a
quantum circuit simulator but by the standard matrix exponentiations,
and thus the Trotterization was absent.
Figure \ref{fig:N2_SVD} (top panel) shows the potential energy curves
of the full-CI and $k$-uCJ ($k$=1,2,3) ansatzs and the SVD low-rank
approximation by Eq.(\ref{eq:low-rank}).  The result of the
conventional CCSD calculations is also shown for comparison.

The 1-uCJ works well at around the equilibrium bond length, e.g. the
energy deviates from the full-CI by 6 mE$_\text{h}$ at $r$=1.2 \AA,
where the multireference effect is not so significant, but it cannot
describe the dissociation correctly.
The error is rapidly decreased by increasing $k$, e.g.\ less than 1
mE$_\text{h}$ at around the equilibrium bond length and at the
dissociation. It was also observed that the convergence to the full-CI
energy is relatively slow at the intermediate region and the maximum
error is 7 mE$_\text{h}$ with $k$=2, and 3 mE$_\text{h}$ with $k$=3 at
$r$=2.0 \AA
%

The SVD low-rank approximation by Eq.(\ref{eq:low-rank}) was found to
be problematic for this system. Not only the convergence to the
reference full-CI energy is slower than the $k$-uCJ even at the
equilibrium bond length region, but we also found that it becomes
pathologically slow when the multireference effect is important.
In the first place, to describe the dissociation correctly by the
low-rank approximation, we needed to optimize the reference T$_2$
amplitude by more expensive ansatz, e.g.\ the uCCGD, that can
correctly describe the dissociation.
The SVD low-rank approximation is not suited for this kind of purpose.

The $k$-uCJ ansatz has similarity to the $k$-UpCCGSD ansatz since both
ansatzs take a product of the single exponential operators uCJ or
UpCCGSD, and the number of parameters grows only quadratically ${\cal
  O}(N^2)$ with the number of orbitals $N$ if $k$ is constant, and
thus is expected to be expressed by a quantum circuit with linear
${\cal O}(N)$ depth.
Figure~\ref{fig:N2_UpCCGSD} (bottom panel) shows the potential
energy curves obtained with $k$-uCJ and $k$-UpCCGSD ansatzs.
The result of 1-uCJ with AGP ansatz, in which the exponential operator of
1-uCJ was applied to the AGP reference wavefunction and all the
parameters including the AGP itself were simultaneously optimize,
is also shown for comparison

The CJ exponential operator was originally combined with the
anti-symmetric geminal power (AGP) reference wavefunction in the
original CJ-AGP ansatz,\cite{neuscamman2013neuscamman-1} which are
expected to serve complementary roles to describe the electron
correlation and can give better description than that with the HF
reference. In fact, its unitary variant, 1-uCJ-AGP ansatz, well
reproduced the full-CI results both at equilibrium bond length region
and at dissociation region with less than 5 mE$_\text{h}$ error. It should be
noted that relatively large error was observed at the intermediate
region, e.g.\ $\sim$ 25 mE$_\text{h}$ error at $r$=1.7 \AA, though the CJ-AGP
ansatz almost perfectly reproduced the full-CI result for whole the
bond stretching ($r$=1.0–1.8) in the original work by Neuscamman. The
unitarization could slightly weaken the flexibility of the CJ ansatz.

The $k$-UpCCGSD ansatz with $k=1$ is not adequate for describing the
PEC, and the error is rapidly decreased by increasing $k$,
while the convergence to the full-CI energy is bit slower than the
$k$-CJ in particular at around the equilibrium bond length and at the
dissociation.
Again, the energies were computed based on the original definitions
written in the second quantized operator, and there is no guarantee
that those can also be reproduced efficiently by the quantum gates on
quantum circuits if the cluster operators are not commutative. This
point will be examined in the next subsection.

\subsection{\label{subsec:results_circ}
Variational energies with the Trotter splitting ansatzs}

\begin{figure}[b]
\captionsetup{justification=raggedright}
\includegraphics[width=8.5cm]{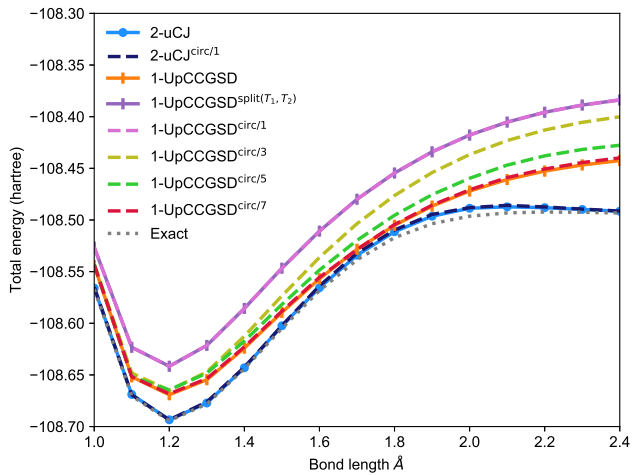}
\caption{\label{fig:N2_circuit} Potential energy curves for the N$_2$
  molecule for the different ansatzs using the STO-6G basis set where
  the four lowest orbitals are frozen, i.e.\ 6 electrons are
  distributed to the other 6 orbitals. The superscript circ/$n$
  denotes a variant of the original ansatz where the Pauli operator
  products are split by $n$-times Trotter steps.}
\end{figure}

In the previous subsection, all the energies were evaluated by using
matrix exponentiation as provided by the definitions written in the
second quantization. As noted above, however, most of the exponential
ansatzs, except for the $k$-uCJ, require the Trotter decomposition
to be accurately computed on quantum circuits. Here, instead of
performing the time consuming many Trotter steps to reproduce the
original ansatzs, we adopted the alternative ansatzs that are defined
by the single or $n$-times Trotter steps for the summation over the
Pauli operator products produced by the Jordan-Wigner transformation
of the original ansatzs.

Figure \ref{fig:N2_circuit} shows the potential energy curves obtained
with the 1-UpCCGSD\circn{n} and $k$-uCJ\circn{n} ansatzs. As expected,
the $k$-uCJ\circn{1}\ reproduced the results of the original $k$-uCJ
ansatz.
However, it is somewhat surprising that the potential energy curve of
the 1-UpCCGSD\circn{1}\ was much different from that of the original
1-UpCCGSD ansatz, in fact the 1-UpCCGSD\circn{1}\ is less powerful
than the 1-UpCCGSD ansatz.
To find a clue to this discrepancies, we performed another variant of
the 1-UpCCGSD in which the singles and doubles operators were split as
$e^{\hat{T}_1 - \hat{T}^\da_1} e^{\hat{T}_2 - \hat{T}^\da_2}$, instead
of $e^{(\hat{T}_1 + \hat{T}_2) - (\hat{T}_1 + \hat{T}_2)^\da}$,
hereafter denoted by 1-UpCCGSD$^{\text{split-}(T_1,T_2)}$.
Interestingly, the potential energy curve of the
UpCCGSD$^{\text{split-}(T_1,T_2)}$ is nearly identical to that of the
1-UpCCGSD\circn{1}. It clearly suggests that the large discrepancies
between the 1-UpCCGSD and 1-UpCCGSD\circn{1} is mainly caused by
splitting the singles and doubles cluster operators, which should be
caused accompanying the splitting of the products of the Pauli
operators in the single Trotter step approximation.
As shown, the 1-UpCCGSD\circn{n} is converged to the result of the
orginal 1-UpCCGSD by increasing the number of the Trotter steps as
$n$=1,3,5.

\section{\label{sec:summary}Summary}
In this study, we present an efficient ${\cal O}(N^2)$-parameter
ansatz for the VQE algorithm, named $k$-uCJ ansatz, which is related
to a tensor decomposition of the coupled cluster
amplitudes,\cite{motta2018chan} specifically generalized CC, and
should be converged to the uCCGD accuracy by increasing the number of
terms, $k$. Each term of the expansion is identical to the unitary
variant of the cluster Jastrow (CJ) ansatz introduced by
Neuscamman,\cite{neuscamman2013neuscamman-1} though the Jastrow
parameters are restricted to pure imaginary by the unitarization and
the physics captured by the ansatz can be different from the CJ
ansatz. It was confirmed that the $k$-CJ ansatz rapidly converged to
the full-CI energy with increasing the $k$, and the chemical accuracy
can be obtained with $k$=3 for the triple bond dissociation of the
N$_2$ molecule.

The accuracy of the exponential ansatzs were examined not only with
their original definition but also with their `hardware-friendly'
variants, in which all the Pauli products generated by the
Jordan-Wigner transformation are split by a few, typically single,
Trotter steps.
Interestingly, the single Trotter step variants gave nearly identical
potential energy curves to the original single exponential ansatz
except for the $k$-UpCCGSD\circn{1}, which is considerably less
accurate than the original $k$-UpCCGSD.

One of the advantages of the $k$-uCJ ansatz for quantum computation is
that it does not need the Trotterization to implement on a quantum
circuit because the cluster operators in each exponent are all
commutative.
The $k$-uCJ ansatz is therefore a good candidate for an efficient
quantum computation with the near-term quantum computers.
It should be noted that one drawback of the use of a product of
exponential operator ansatz is that the parameter optimization is often
trapped by the local minimum, as has been already reported by
Lee.\cite{lee2019whaley}
We also found that the use of the anti-symmetrized geminal power (AGP)
reference function can improve the description of the $k$-uCJ, as the
CJAGP ansatz does, thus to improve the reference function should be
another issue to be explored.

In the NISQ era,\cite{preskill2018preskill} quantum computers can
treat a few hundred orbitals at most, which is not quite large enough
for describing the dynamical correlation in most cases. A
straightforward approach to this problem is to adopt the active space
model and use the VQE as an alternative to the full-CI in the active
space, as done in the DMRG-CASSCF and subsequent dynamical correlation
theories.\cite{zgid2008nooijen, zgid2008nooijen-3, ghosh2008chan,
  kurashige2009yanai, yanai2010chan, kurashige2011yanai,
  saitow2013yanai, saitow2015yanai} In that case, the cumulant
approximation to the high-order reduced density matrices should be
necessary to reduce the required times of the
measurements.\cite{zgid2009chan, kurashige2014yanai-2,
  takeshita2019mcclean}
Another approach is to cast the dynamical correlation into the active
space Hamiltonian by similarity transformations in the first or second
quantization, e.g.\ Ref~[\onlinecite{yanai2012shiozaki,
    watson2016chan}] and Ref~[\onlinecite{boys1969handy,
    nooijen1998bartlett, hino2002ten-no, cohen2019alavi}],
respectively, and many others though it is not possible to list them
all here.
The $k$-uCJ ansatz is particularly suitable for this approach because
the similarity transformed Hamiltonians usually possess more than
two-body interactions and it is straightforward to extend the $k$-uCJ
ansatz to more than two-body Jastrow factor forms.

\section{\label{sec:computational}Computational Details}
An in-house code written in Python 3 was used to perform all the
calculations. The HF canonical orbitals and molecular integrals were
generated by the PySCF library,\cite{PYSCF} and the fermionic algebra
for the exponential ansatz in the Fock space and the mapping to a
qubit representation by the Jordan-Wigner transformation were handled
by using the OpenFermion library.\cite{mcclean2017openfermion} The
quantum circuit simulations were performed by the Qulacs
library\cite{Qulacs}.

%
\bibliographystyle{achemso}
\bibliography{zotero}                              

\onecolumngrid

\section*{\label{sec:appendix} Appendix: Jastrow-type decomposition of the Hamiltonian}
\begin{figure*}[b]
\captionsetup{justification=raggedright}
    %
    \begin{minipage}[t]{0.45\hsize}
      \subcaption{$k$ = 2}
      \vspace{-2mm}
      \centering
      \includegraphics[width=7cm]{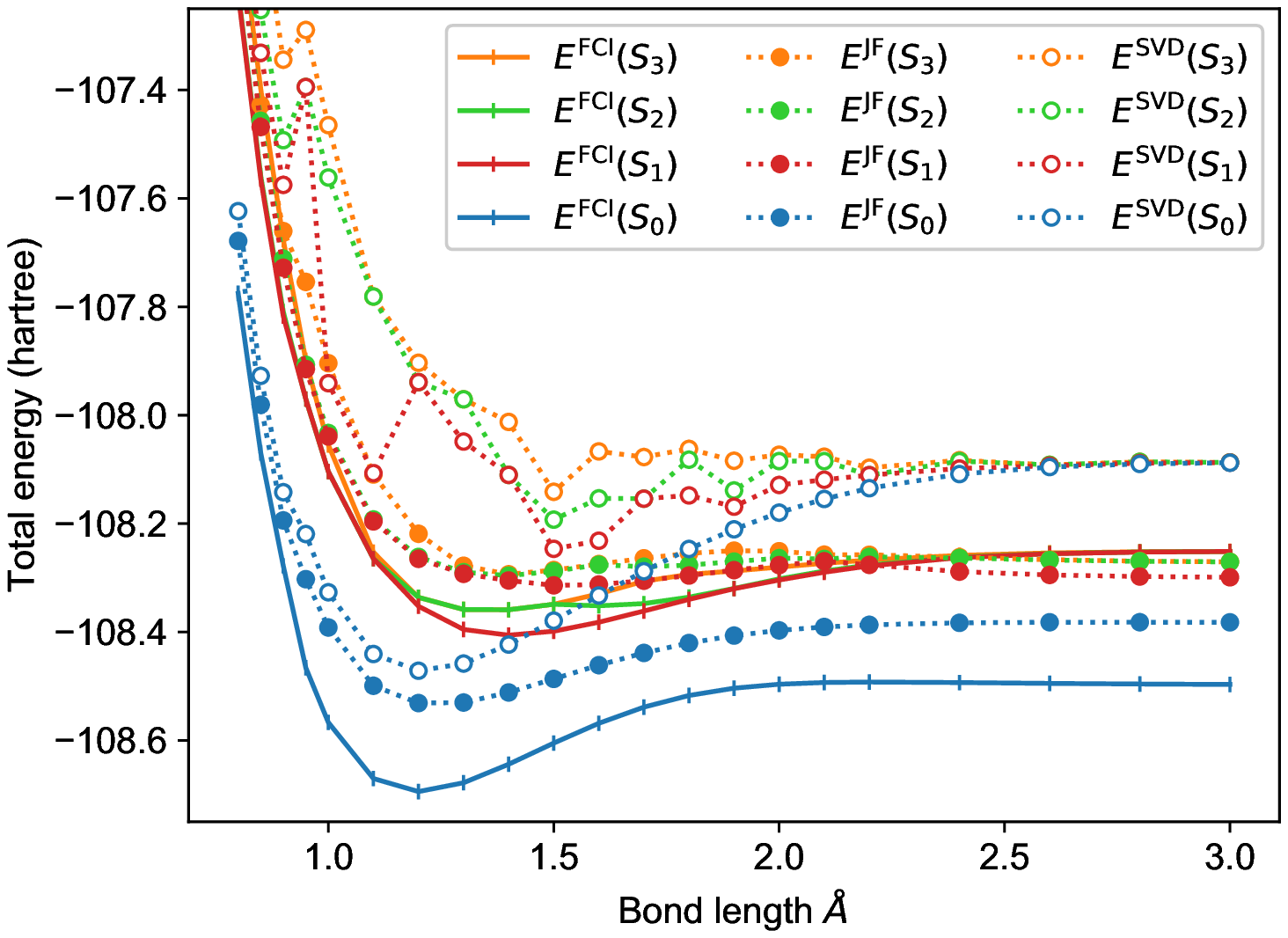}
      \label{fig:N2_PEC_L-2}
    \end{minipage} 
    %
    \begin{minipage}[t]{0.45\hsize}
      \subcaption{$k$ = 4}
      \vspace{-2mm}
      \centering
      \includegraphics[width=7cm]{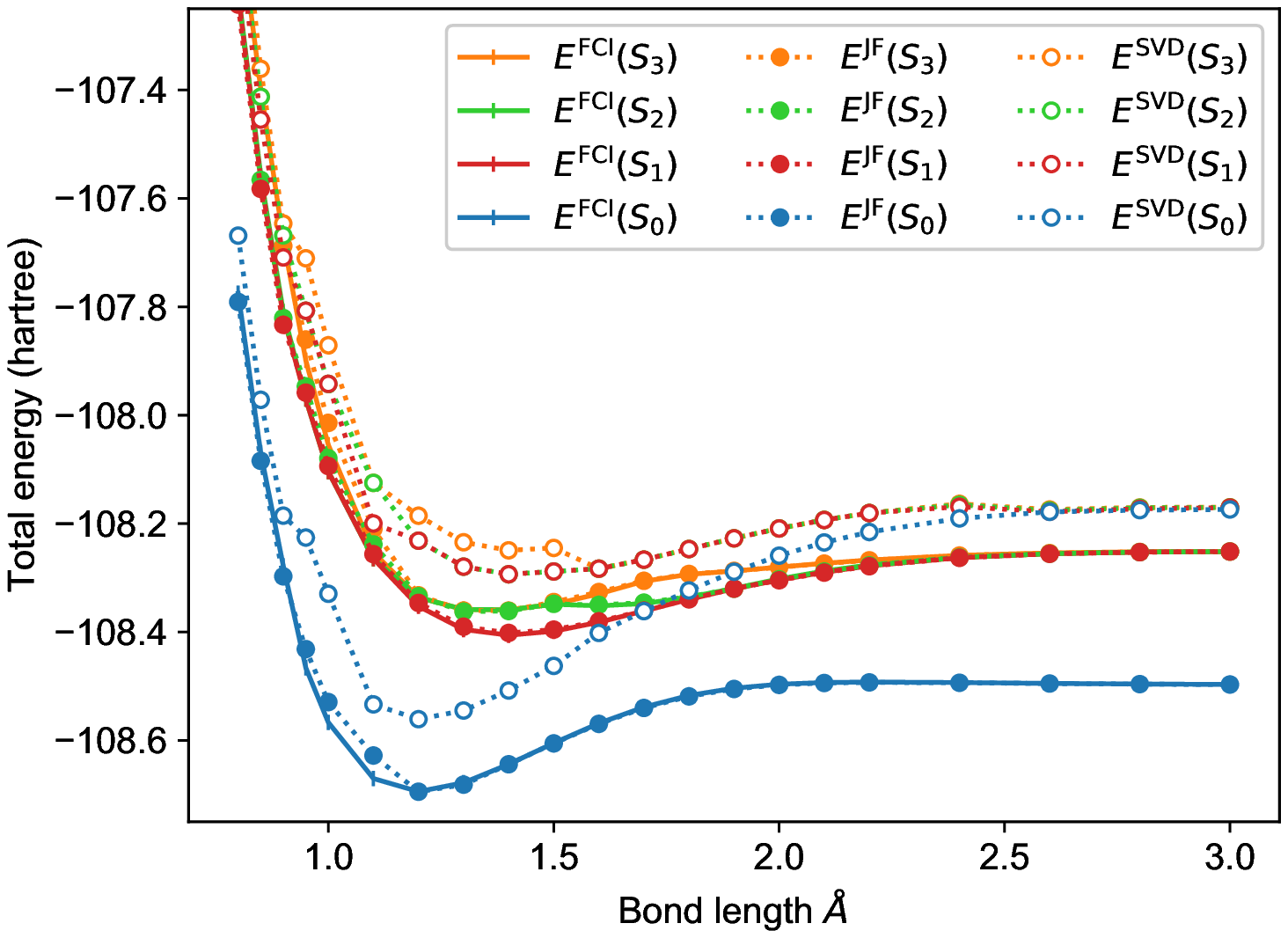}
      \label{fig:N2_PEC_L-4}
    \end{minipage}
    \\
    \begin{minipage}[b]{0.45\hsize}
      \subcaption{$k$ = 8}
      \vspace{-2mm}
      \centering
      \includegraphics[width=7cm]{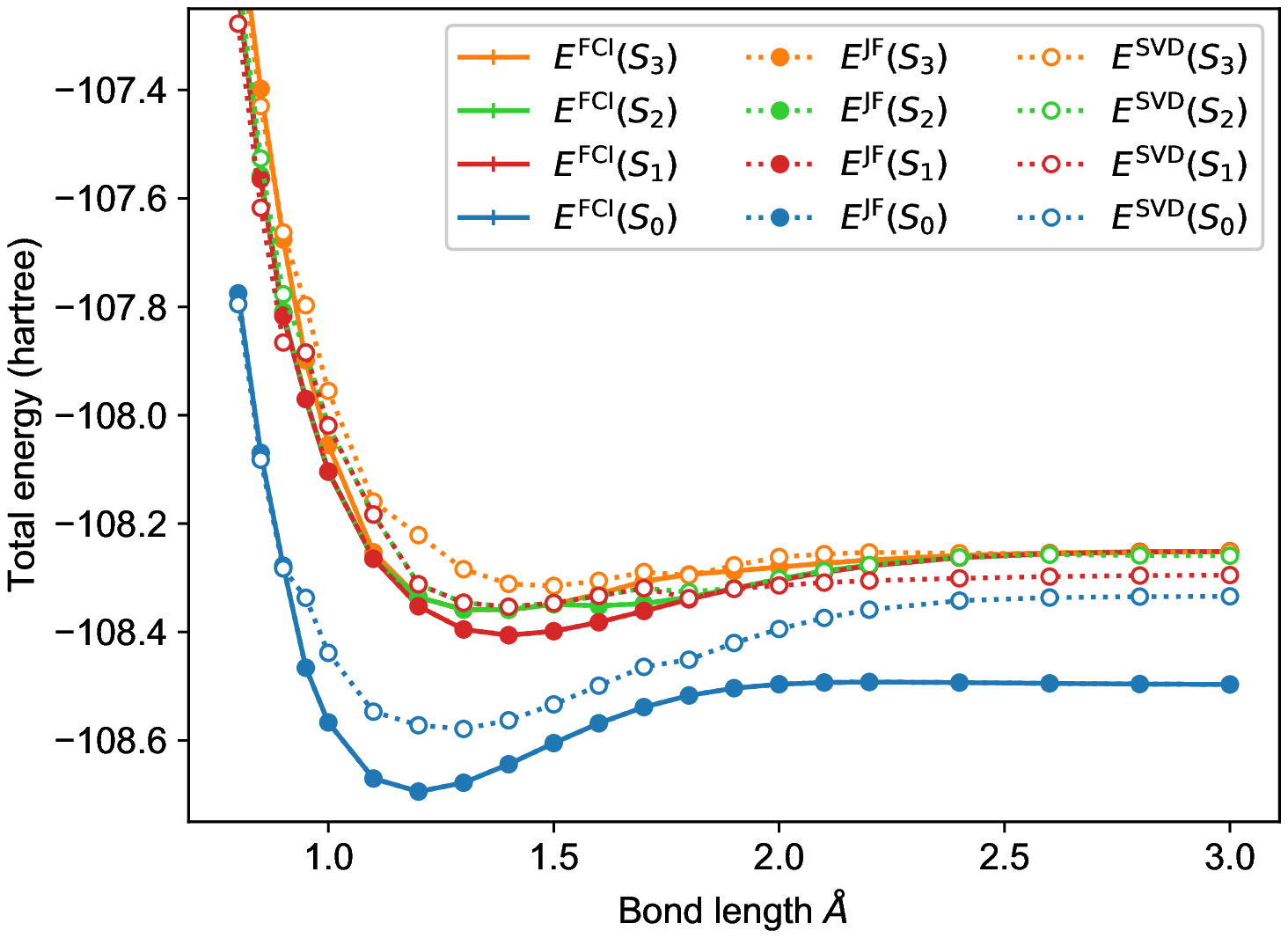}
      \label{fig:N2_PEC_L-8}
    \end{minipage} 
    %
    \begin{minipage}[b]{0.45\hsize}
      \subcaption{$k$ = 16}
      \vspace{-2mm}
      \centering
      \includegraphics[width=7cm]{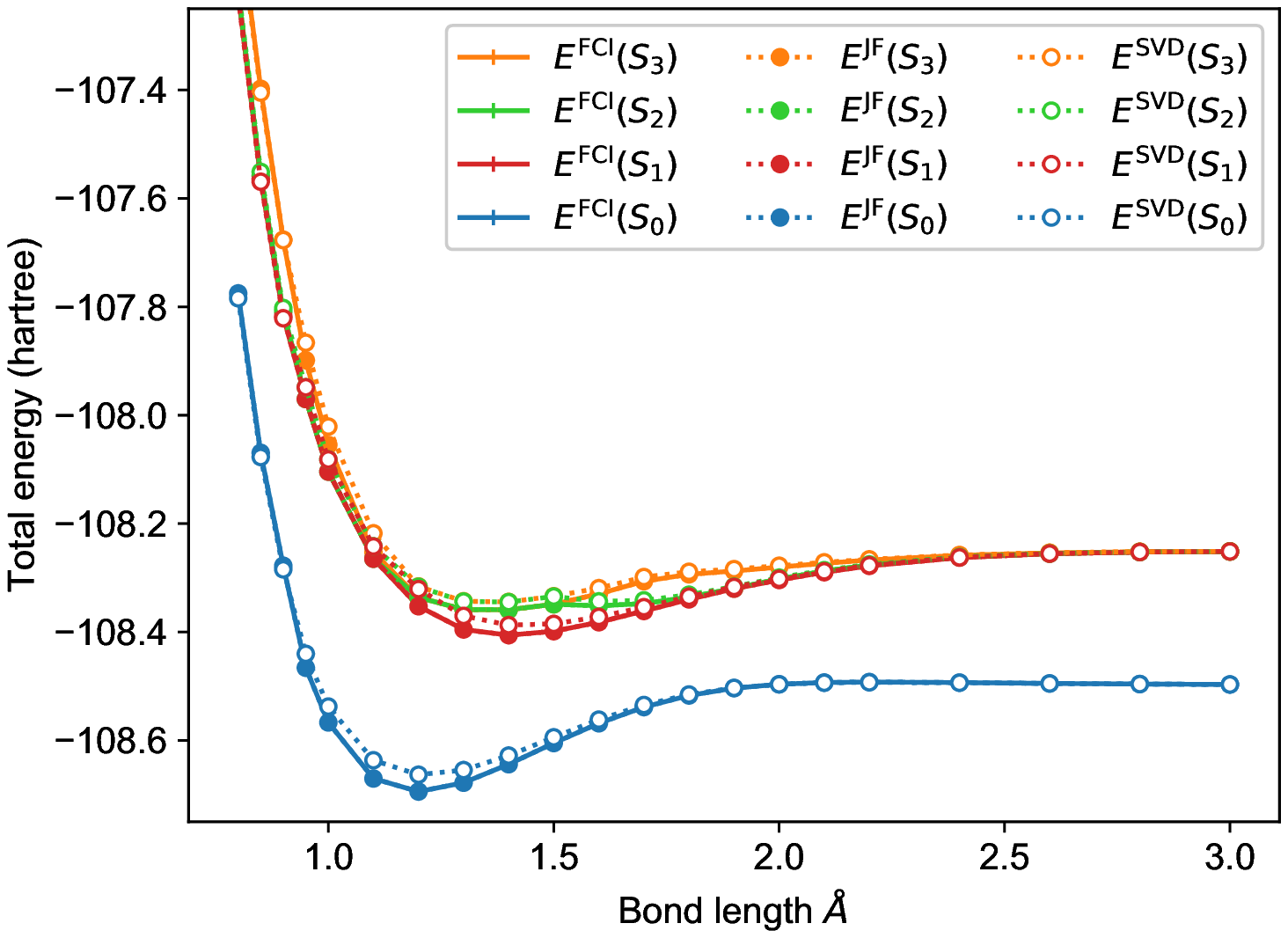}
      \label{fig:N2_PEC_L-16}
    \end{minipage}
    %
    \caption{Potential energy curves for the low-lying singlet states
      of the N$_2$ molecule using the STO-6G basis set where the four
      lowest orbitals are frozen, i.e.\ 6 electrons are distributed to
      the other 6 orbitals. $E^{\text{JF}}$ and $E^{\text{SVD}}$ were
      obtained by diagonalizing the approximated Hamiltonian given by
      Eq.(\ref{eq:H_JF}) and Eq.(\ref{eq:H_SVD}), respectively.}
    \label{fig:N2_decomp}
\end{figure*}

%
%
The four-index two-electron replusion integrals (ERI) in {\it ab
  initio} Hamiltonian
$
\hat{H} = \sum_{pq,\si} f_{pq} a^\da_{p\si} a_{q\si}
        + \sum_{pqrs,\si\ta} h_{pqrs} a^\da_{p\si} a_{q\si} a^\da_{r\ta} a_{s\ta}.
$
%
can also be decomposed by the Jastrow-type parametrization, which is
rather simpler form than that for T$_2$ amplitudes in the previous
subsection \ref{subsec:theory_amp} due to its 8-fold symmetry,
\begin{align}
h_{pq,rs}
&\simeq h^\text{\tiny JF}_{pq,rs}
=\sum_{x} \sum_{kl} U^x_{pj} U^x_{qj} J^x_{jl} U^x_{rl} U^x_{sl}\label{eq:H_JF}
\\
\hat{H}
&\simeq \hat{H}^\text{\tiny JF}
= \sum_{pq,\si} f_{pq} a^\da_{p\si} a_{q\si}
+ \sum^k_{x=1} \sum_{jl,\si\ta} J^x_{jl} \atil^{x\da}_{j\si} \atil^x_{j\si} \atil^{x\da}_{l\ta} \atil^x_{l\ta}%
 = \sum_{pq,\si} f_{pq} a^\da_{p\si} a_{q\si}
+ \sum^k_{x=1} e^{-\hat{K}^{x}} \hat{J}^{x} e^{\hat{K}^{x}}
\end{align}
where $\atil^{x\da}_{j\si} = \sum_p a^\da_p U^x_{pj}$ and $K^{x}$ and
$J^{x}$ are real matrices as in
Eq.(\ref{eq:matJ}),(\ref{eq:matK}). This Jastrow-type parametrization
should be compared with the low-rank approximation to Hamiltonian
introduced by Peng and Kowalski\cite{peng2017kowalski} and recently by
Motta and co-workers.\cite{motta2018chan, berry2019babbush}
\begin{align}
  h_{pq,rs}
  \simeq h^\text{\tiny SVD}_{pq,rs}
  &= \sum^k_{x=1} \Big(\sum_i U^{x}_{pj} \la^{x}_{j} U^{x}_{qj}\Big)
                  \Big(\sum_j U^{x}_{rl} \la^{x}_{l} U^{x}_{sl}\Big)
  =\sum_{x} \sum_{kl} U^x_{pj} U^x_{qj} \la^{x}_{j}
           \la^{x}_{l} U^x_{rl} U^x_{sl}
,\label{eq:H_SVD}
\end{align}
While full-rank matrices $J^{x}$ are not separable for the indices $j$
and $l$ and thus may not be suitable for reducing the cost of the
atomic-orbital integral transformation in classical algorithms, but
the Jastrow-type parametrization is again more flexible than the
low-rank approximation with the same number of terms, $k$, as shown in
Figure \ref{fig:N2_decomp}. The parameters were determined by
minimizing the error $\Delta = |{\bf h}-{\bf h}^\text{\tiny JF}|$.
This should be advantageous when it is used for real-time propagation
by $e^{-i\hat{H}t}$ or the evaluation of the energy expectation values
for VQE on quantum computers,
e.g.\ Ref[\onlinecite{huggins2019babbush}] and references therein.

\end{document}